\documentstyle[12pt]{article}
\begin{document}
\begin{titlepage}
\hskip 12cm \vbox{\hbox{BUDKERINP/98-56} \hbox{July 1998}}
\vskip 0.3cm
\centerline{\bf NEXT-TO-LEADING BFKL$^{~\ast}$}
\centerline{  V.S. Fadin$^{\dagger}$}
\centerline{\sl Budker Institute for Nuclear Physics,}
\centerline{\sl Novosibirsk State University, 630090
Novosibirsk, Russia}

\begin{abstract}
The representation of
the total cross section at  high energy $\sqrt s$
in the next-to-leading $\ln s$ approximation is given
with  definition of the impact factors and explicit expression for
the BFKL kernel. The estimate of the Pomeron intercept and the
next-to-leading contributions to anomalous dimensions near
$j=1$ are presented.
\end{abstract}
\vskip .5cm
\hrule
\vskip.3cm

\noindent

\noindent
$^{\ast}${Talk given at the International Conference "DIS98",  April 4-8, Brussels, Belgium, 1998}
\vfill
$ \begin{array}{ll}
^{\dagger}\mbox{{\it email address:}} &
 \mbox{FADIN~@INP.NSK.SU}\\
\end{array}
$
\vfill
\vskip .1cm
\vfill
\end{titlepage}
\eject
\textheight 210mm \topmargin 2mm \baselineskip=24pt

\vskip 0.5cm

At large c.m.s. energy  $\sqrt s$ in the leading logarithmic approximation
(LLA)
the total cross-section $\sigma (s)$ for the high energy scattering
of colourless particles $A,B$ can be presented \cite{FKL} in terms of
their impact factors $\Phi _i(\overrightarrow{q_i})$ and
the Green function $G(s; \vec q_1,\vec q_2 )$
for the reggeized gluon scattering at zero momentum transfer.
Let us consider the impact factors as
(not normalized) wave functions of the $t$-channel two-particle states,
denote these states $\left| A\right\rangle $ and $\left| B\right\rangle$
and use the complete set of the states  $\left| \vec q\right\rangle $
in the transverse momentum space with
the properties $\left\langle \vec q_1\right. \left| \vec q_2\right\rangle =
\delta(\vec q_1-\vec q_2)\,,\,\, \stackrel{\wedge }{q^2}\left| \vec q_i\right\rangle =
{q_i^2}\left| \vec q_i\right\rangle\,,$  so that
$\Phi _A(\vec q)\,\,={2\pi q^2}\left\langle
\vec q\right. \left| A\right\rangle\,. $
In these denotations
\begin{equation}
\sigma (s)=
\left\langle A\right| \,
\stackrel{\wedge }{G}(s)\,\,
\left| B\right\rangle \,, \,\,\, \stackrel{\wedge }{G}(s)
=\left( \frac s{s_0}\right)^{\stackrel{\wedge }{K}}\,
\label{I10}
\end{equation}
with the kernel
\begin{equation}
\left\langle \vec q_1\right| \stackrel{\wedge }{K}\left|
\vec q_2\right\rangle =2\,\omega (q_1)\,\delta (%
\vec q_1-\vec q_2)+K_r(\vec q_1,%
\vec q_2)  \label{I12}
\end{equation}
which is expressed in terms of the gluon Regge trajectory $\omega (q)$ and
the integral
kernel $K_r(\vec q_1,\vec q_2)$, related with the
real particle production. Taking separately $\omega (q)$ and
$K_r(\vec q_1,\vec q_2)$ contain the infrared divergencies; after their
cancellation  the kernel averaged  over the angle between the
momenta $%
\vec q_1$ and $\vec q_2$  can be presented  in LLA as
\begin{equation}
\overline{\left\langle \vec q_1\right| \stackrel{\wedge }{K^B}\left| \vec q_2\right\rangle }=\frac{\alpha _s(\mu^2) N_c}{\pi ^2%
}\int \frac{dq^2}{|q_1^{~2}-q^{~2}|}\left( \delta (q^2-q_2^2)-2\frac{\min
(q_1^2,q^2)}{(q_1^2+q^2)}\delta (q_1^2-q_2^2)\right).
\label{IB}
\end{equation}
Remind, that the dependence from the energy scale $s_0$ as well as from
the argument $\mu^2$  of the coupling constant is beyond the LLA accuracy.

The program of calculating next-to-leading corrections to the BFKL equation
was formulated several years ago \cite{LF}. It was shown, that in the
next-to-leading logarithmic approximation (NLLA), due to the gluon
Reggeization,
the form (\ref{I10}) of the cross section,
as well as the representation (\ref{I12}) for the kernel remain unchanged,
but the trajectory $\omega(q)$ should be taken with the two-loop
correction (it was obtained in \cite{FFK}) and the integral kernel $K_r(\overrightarrow{q_1%
},\overrightarrow{q_2})$ - with  one-loop accuracy.
The one-loop correction to the integral kernel  is obtained as a sum of two contributions. The
first one is related with the one-loop virtual correction to the one-gluon
production cross-section \cite{FL} and the second one is determined by the Born cross-sections for
production of two gluons \cite{FL96} and quark-antiquark pair \cite{CCH}.

For the total NLLA kernel it was obtained
\cite{FL98}:
\[
\overline{\left\langle \vec q_1\right| \stackrel{\wedge }{K}%
\left| \vec q_2\right\rangle }=\frac{\alpha _s(\mu ^2)N_c}{\pi ^2%
}\int dq^2\frac 1{|q_1^{~2}-q^{~2}|}\left( \delta (q^2-q_2^2)-2\frac{\min
(q_1^2,q^2)}{(q_1^2+q^2)}\delta (q_1^2-q_2^2)\right)
\]
\[
\times \left[ 1-\frac{\alpha _s(\mu ^2)N_c}{4\pi }\left( \left( \frac{11}3-%
\frac{2n_f}{3N_c}\right) \ln {\left( \frac{|q_1^{~2}-q^{~2}|^2}{\max
(q_1^2,q^2)\mu ^2}\right) }-\left( \frac{67}9-\frac{\pi ^2}3-\frac{10}9\frac{%
n_f}{N_c}\right) \right) \right]
\]
\[
-\frac{\alpha _s^2(\mu ^2)N_c^2}{4\pi ^3}\left[ \frac 1{32}\left( 1+\frac{n_f%
}{N_c^3}\right) \left( \frac 2{{q_2}^2}+\frac 2{{q_1}^2}+(\frac 1{{q_2}%
^2}-\frac 1{{q_1}^2})\ln \frac{{q_1}^2}{{q_2}^2}\right)
+\frac{\left( \ln {(q_1^2/q_2^2)}\right)^2} {|q_1^2-q_2^2|} \right.
\]
\[
\left.  +\left( 3+(1+\frac{n_f}{N^3})\left( \frac 34-\frac{(q_1^2+q_2^2)^2}{%
32q_1^2q_2^2}\right) \right) \int_0^\infty \frac{dx}{q_1^2+x^2q_2^2}\ln
\left| \frac{1+x}{1-x}\right| \right] -\frac 1{q_2^2+q_1^2}\left( \frac{\pi ^2}3
\right.
\]
\begin{equation}
\left.-4L(\min (\frac{q_1^2}{q_2^2},%
\frac{q_2^2}{q_1^2}))\right)
+\frac{\alpha _s^2(\mu ^2)N_c^2}{4\pi ^3}\left( 6\zeta (3)-\frac{5\pi ^2}{%
12}\left( \frac{11}3-\frac{2n_f}{3N_c}\right) \right) \delta (q_1^2-q_2^2)\,,
\label{I18}
\end{equation}
where
\begin{equation}
L(z)=\int_0^z\frac{dt}t\ln (1-t)\,,\,\,\zeta (n)=\sum_{k=1}^\infty k^{-n}.
\label{I19}
\end{equation}
In NLLA the impact factors  in (\ref{I10}) depend
on the energy scale $s_0$:
\[
\Phi_P({\vec q}_R;s_0)= \frac{\sqrt{N_c^2-1}}{2\pi s}\int d s_{PR}
I_{PR}\sigma_{PR}^{(s_0)}(s_{PR})\theta(s_{\Lambda}-s_{PR})
\]
\begin{equation}
-\int \frac{d \vec q}{(2\pi)^{D-1}}
\Phi^{(B)}_P(\vec q)\frac{g^2N_c \vec q_R^{~2}}{\vec q^{~2}
(\vec q_R-\vec q)^2}\ln \left(\frac{s_{\Lambda}^2}{(\vec q_R-
\vec q)^2s_0}
\right).
\label{I7}
\end{equation}
Here $D$ is the space-time dimension, $\sigma_{PR}^{(s_0)}(s_{PR})$ is the total cross section of the
particle-Reggeon scattering at the c.m.s. energy $\sqrt {s_{PR}}$
averaged over colours of the Reggeon   with the one-loop corrections
calculated for  case when the energy scale
in the Regge factors is equal $s_0$;  $I_{PR}$ is the invariant flux,
$I_{PR}=\sqrt{(s_{PR}-p_P^2+\vec q_R^{~2})^2+4p_P^2\vec q_R^{~2}}\, ,$
and $\Phi^{(B)}_P(\vec q_R)$ is the Born (LLA) value of the impact
factor. The right hand part of the above equation is assumed to be taken in
the limit $s_\Lambda \rightarrow \infty$,  so that the dependence
from $s_\Lambda$ disappears due to the factorization properties
of the Reggeon vertices in the regions of strongly ordered
rapidities of the produced particles.

In Ref. \cite{FL98}  the energy scale $s_0$ was taken equal $q_1q_2$
 which
is natural from the point of view of the
Watson-Sommerfeld representation for high energy scattering amplitudes. It was
pointed out in Ref. \cite{FL98}, that the change of the energy scale leads
generally to the corresponding modification of the impact factors and the
BFKL equation for the Green function $G$,  but the physical results are
not changed.
Since with the NLLA accuracy
\[
\left( \frac s{s_0}\right)^{\stackrel{\wedge }{K }} = s^{\stackrel{\wedge }{K }}(1-{\stackrel{\wedge }{K^B }} \ln {s_0}) =\int d\vec q_1 \int d\vec q_2
\left(1+ \ln {\left(\frac{f_1(\stackrel{\wedge }{q^2})}{s_0}\right)}
\frac{\stackrel{\wedge }{K^B }}2\right)
\]
\begin{equation}
\times
\left|\vec q_1\right\rangle\left\langle \vec q_1\right|
\left(\frac s{\sqrt{f_1({q_1^2})f_2({q_2^2})}}\right)^{\stackrel{\wedge }{K }}
\left|\vec q_2\right\rangle\left\langle \vec q_2\right|
\left(1+\frac{\stackrel{\wedge }{K^B }}2
\ln {\left(\frac{f_1(\stackrel{\wedge }{q^2})}{s_0}\right)}
\right)\,,
\label{I13}
\end{equation}
where $f_1$ and $f_2$ are some functions,
we obtain that at the transition from the scale $s_0$ to any factorizable
scale $\sqrt{f_1({q_1^2})f_2({q_2^2})}$ we can keep the kernel unchanged,
changing the impact factors to
\begin{equation}
{\Phi} _P(\vec q_i;f_i(q_i^2))={\Phi} _P(\vec q_i;s_0)+\frac{1}{2}
\int d\vec q\,{\Phi}^{B} _P(\vec q)\ln \left(\frac{f_i(q^2)}{s_0}\right)
K^B(\vec q, \vec q_i)\frac{q_i^2}{q^2}.
\label{I17}
\end{equation}

The action of the kernel (\ref{I18})
on the eigenfunctions $q_2^{2(\gamma -1)}$ of the Born kernel gives us
\cite{FL98}
\begin{equation}
\int d\vec q_2\,\,\,K(\vec q_1,\vec q_2)\left(
\frac{q_2^2}{q_1^2}\right) ^{\gamma -1}=\frac{\alpha _s(q_1^2)\,N_c\,}\pi
\left( \chi^{(B)} (\gamma )+\frac{\alpha _s(q_1^2)N_c}{\pi }
\chi^{(1)} (\gamma )%
\right) \,,\, \label{I20}
\end{equation}
were  $\chi^{(B)} (\gamma )=2\psi (1)-\psi (\gamma )-\psi (1-\gamma )\,$
is
proportional to the eigenvalue of the Born kernel, $\psi
(\gamma )=\Gamma ^{\prime }(\gamma )/\Gamma (\gamma )\,,  $
 and  the correction $\chi^{(1)} (\gamma )$ is:
\[
\chi^{(1)} (\gamma )=-\frac{1}{4}\left[ \left( \frac{11}3-\frac{2n_f}{3N_c}\right) \frac
12\left( \chi ^2(\gamma )-\psi ^{\prime }(\gamma )+\psi ^{\prime }(1-\gamma
)\right) \right.
\]
\[
\left. -6\zeta (3)+\frac{\pi ^2\cos(\pi \gamma )}{\sin^2(\pi \gamma
)(1-2\gamma )}\left( 3+\left( 1+\frac{n_f}{N_c^3}\right) \frac{2+3\gamma
(1-\gamma )}{(3-2\gamma )(1+2\gamma )}\right) \right.
\]
\begin{equation}
\left.-\left( \frac{67}9-\frac{\pi ^2}3-\frac{10}9\frac{n_f}{N_c}\right)
\chi (\gamma ) -\psi ^{\prime \prime }(\gamma )-\psi ^{\prime \prime }(1-\gamma )-%
\frac{\pi ^3}{\sin (\pi \gamma )}+4\phi (\gamma )\right] \,  \label{I22}
\end{equation}
with
\[
\phi (\gamma )=-\int_0^1\frac{dx}{1+x}\left( x^{\gamma -1}+x^{-\gamma
}\right) \int_x^1\frac{dt}t\ln (1-t)
\]
\begin{equation}
=\sum_{n=0}^\infty (-1)^n\left[ \frac{\psi (n+1+\gamma )-\psi (1)}{(n+\gamma
)^2}+\frac{\psi (n+2-\gamma )-\psi (1)}{(n+1-\gamma )^2}\right] \,.
\label{I23}
\end{equation}

For the relative correction $r(\gamma )$  defined as
 $\chi^{(1)} (\gamma )=-r(\gamma )\chi^{(B)} (\gamma )\,$
in the symmetrical point $\gamma =1/2$, corresponding to the
Pomeron singularity  we have
$r( 1/2) \simeq 6,46 +0.05\frac{n_f}{N_c}+2.66\frac{n_f}{%
N_c^3}$, i.e., the correction is large.
In some sense, the large value of the correction is natural
and is a consequence
of the large value of the  Born intercept $\omega_P
^B=4N_c\ln 2\alpha _s(q^2)/\pi\,$. If we express the corrected intercept $\omega_P$
in terms of the Born one,
we obtain
\begin{equation}
\omega_P = \omega_P^B(1-\frac{r\left( \frac 12\right)}{4\ln 2}\omega_P^B)
\simeq \omega_P^B(1-2.4\omega_P^B).
\label{I25}
\end{equation}
The coefficient 2.4 does not look very large. Moreover, it corresponds to
the rapidity interval where correlations become important in
the hadron production processes. Nevertheless, if we take
(\ref{I25}) for $\alpha_s(q^2)=0.15$ we obtain $\omega_P\simeq 0.07$
that is too small. But it is necessary to realize that, firstly,
the estimate (\ref{I25}) is quite straightforward and does not take into account
neither the influence of the running coupling  on the eigenfunctions nor the
nonperturbative effects \cite{L}; secondly, the value of the correction
strongly
depends on its representation.
For example, if one
takes into account the next-to-leading correction by
the corresponding increase of the argument of the running QCD coupling
constant, $\omega_P$ at $\alpha _s(q^2)=0.15$ turns out to be only two
times smaller, than its Born value.

The results obtained for the BFKL kernel can be applied for the calculation
of anomalous dimensions
of the local twist-2 operators near point $\omega =J-1=0$, which are
determined \cite{FL98} from the
solution of the equation
\begin{equation}
\omega =\frac{\alpha _sN_c}\pi \left( \chi^{(B)} (\gamma )
+\frac{\alpha _s N_c}{\pi }(\chi^{(1)}(\gamma )-2\chi^{(B)}(\gamma )
(\chi^{(B)}(\gamma ))^{\prime})\right).  \label{I30}
\end{equation}
For the low orders of the perturbation theory the solution of (\ref{I30})
reproduces the
known results and gives  the higher loop correction \cite{FL98}
for $\omega \rightarrow
0$:
\[
 \gamma \simeq \frac{\alpha _s\,N_c}\pi (\frac 1\omega
 -\frac{11}{12}-\frac{n_f}{%
 6N_c^3})-\left( \frac{\alpha _s\,}\pi \right)
 ^2\frac{n_f\,N_c}{6\,\omega }%
 (\frac 53+\frac{13}{6N_c^2})
 \]
 \begin{equation}
 -\frac 1{4\omega ^2}\left( \frac{\alpha _sN_c}\pi \right) ^3
 \left( \frac{395}{27}-2\zeta (3)-%
 \frac{11}3\frac{\pi ^2}6+\frac{n_f}{N_c^3}(\frac{71}{27}-\frac{\pi
 ^2}%
 9)\right) \,\,.
\label{I31}
\end{equation}

The results obtained in \cite{FL98} for the BFKL kernel and the anomalous
dimensions were confirmed in \cite{CC}.
\vskip 1.5cm
{\bf Acknowledgment}

I thank the Organizing Committee for the financial
support of my participation in the Workshop. The presented results were
obtained in researches supported by the INTAS and
RFFI grants.

\end{document}